\documentclass[twocolumn,showpacs,aps,prb]{revtex4}% Physical Review Letters
\usepackage{graphicx}% Include figure files
\usepackage{dcolumn}% Align table columns on decimal point
\usepackage{bm}% bold math

\begin{document}

\title{Fine-structure splitting reduction of ionized impurity bound exciton in quantum dot
%Reduction of the exciton fine-structure splitting in semiconductor quantum dots
%induced by ionized impurity centers
%Ionized impurity induced reduction of the exciton fine-structure splitting in semiconductor quantum dots
% Reduction of the exciton fine-structure splitting and abnormal Stark effect in quantum dots induced by impurity centers
}
\author{Dong Xu}
%\email[Electronic address: ]{d-xu03@mails.tsinghua.edu.cn}
\affiliation{%
Department of Physics, Tsinghua University, Beijing 100084, People's Republic of China}
\author{Jia-Lin Zhu}
\email[Author to whom correspondence should be addressed; electronic address:
]{zjl-dmp@tsinghua.edu.cn}
\affiliation{%
Department of Physics, Tsinghua University, Beijing 100084, People's Republic
of China}
\date{\today}% It is always \today, today,

\begin{abstract}
The ground-state energy and fine-structure splitting of ionized shallow donor
impurity-exciton complex in quantum dots are investigated. It is found that
fine-structure splitting could be largely reduced by the off-center ionized
impurities since the anisotropic shape of exciton envelope function is
significantly changed. Anomalous Stark shifts of the ground-state energy and
efficient tuning of the fine-structure splitting by the external electric field
due to the local electric field produced by the ionized impurities are
discussed. The scheme may be useful for the design of the quantum dots-based
entangled-photon source.
\end{abstract}
\maketitle

Semiconductor quantum dots (QDs) have been demonstrated as one of candidates
for the entangled two-photon sources, which make them very attractive for
applications in the fields of quantum teleportation and quantum
computation.~\cite{Akopian,Stevenson} Much efforts, e.g., thermal annealing and
external field
tuning,~\cite{Young,Tartakovskii,Seguin2,Stevenson2,Kowalik,Gerardot} have been
devoted to the reduction of the fine-structure splitting (FSS) of the
intermediate exciton states, since a necessary condition in the proposal for a
QD-based source of polarization entangled photon pairs is that the intermediate
exciton states for the biexciton radiative decay are energetically
degenerate.~\cite{Benson} FSS originated from the anisotropic electron-hole
exchange interaction is directly determined by the anisotropic shape of the
exciton envelope function. Obviously, exciton envelope function may be strongly
influenced by the ionized shallow impurities in QDs, and consequently FSS might
be changed. In this letter, we study the ground-state FSS of ionized
hydrogenetic donor impurity-exciton complex in quantum dots and find that it
could be largely reduced by the ionized hydrogenetic impurities as well as the
external electric field.

%However, the monoexciton ground states in III-V semiconductor QDs are split by
%the anisotropic electron-hole exchange interaction since QDs tend to be
%elongated along the $[\bar{1}10]$ crystal
%axis.~\cite{Gammon,Bayer,Ivchenko,Takagahara,Bester1,Seguin}

Recently, with the advancement of QDs growth and measurement technics, it is
possible to optically probe fine-structures of single magnetic impurity-doped
semiconductor QDs.~\cite{Besombes} In this letter, the ground-state energy and
FSS of ionized hydrogenetic donor impurity-exciton complex in semiconductor QDs
under an external in-plane electric field are investigated, since shallow donor
impurity is common and well studied in III-V semiconductors, e.g., Si-doped
GaAs. The light-hole and spin-orbit-split $J=1/2$ valence bands could be
reasonably neglected in the calculations, since the heavy-hole component is
dominant in the hole ground state of flat InGaAs QDs~\cite{Sheng} and we mainly
focus on the exciton ground states. Thus the exciton state is composed of 4
combinations of the valence band and the conduction band, i.e.,
$|X\rangle=\sum_{m,s}\sum_{r_e,r_h}
\psi_{ms}(r_e,r_h)a^{\dag}_{c_sr_e}a_{v_mr_h}|0\rangle$ where the Wannier
function representation of the creation and annihilation operators is used,
$\psi_{ms}(r_e,r_h)$ is the exciton envelope function, and $m$ and $s$ are the
$z$-component of the angular momentum of the heavy-hole valence band and the
conduction band, respectively. The eigenvalue equation for $\psi_{ms}$ is given
as {\setlength\arraycolsep{2pt}
\begin{eqnarray}\label{eigenvalue equation}
\sum_{m's'r'_er'_h}[H_1&+&V_{\mathrm{ex}}(c_sr_e,v_{m'}r'_h;c_{s'}r'_e,v_mr_h)]\psi_{m's'}(r'_e,r'_h)
\nonumber\\&&=E\psi_{ms}(r_e,r_h),
\end{eqnarray}}
with
\begin{eqnarray}\label{spin-independent Hamiltonian}
H_1=\delta_{r_er'_e}\delta_{r_hr'_h}\delta_{s's}\delta_{m'm}[H_e + H_h
\nonumber\\+e\mathrm{F}\cdot(\mathrm{r}_e-\mathrm{r}_h)-\frac{e^2}{\epsilon|\mathrm{r}_e-\mathrm{r}_h|}
+V_{\mathrm{int}}(\mathrm{r}_h,\mathrm{r}_e,\mathrm{q}_j)],
\end{eqnarray}
where $H_k=p^2_k/2m_k+U_k(r_k)$ $(k=e,h)$ is the single particle Hamiltonian,
$U_h$ ($U_e$) is the confinement potential for the hole (electron),
$\mathrm{F}$ is an external in-plane electric field, and
$V_{\mathrm{int}}(\mathrm{r}_h,\mathrm{r}_e,\mathrm{q}_j)=\sum^{N}_{j=1}(e^2/\epsilon|\mathrm{r}_h-\mathrm{q}_j|-e^2/\epsilon|\mathrm{r}_e-\mathrm{q}_j|)$
is the Coulomb interaction between charge carriers and ionized impurity
centers, $\mathrm{q}_j=(x_j,y_j)$ is the position of the $j$th ionized donor
impurity, $N$ is the total number of ionized donor impurities. Similar to the
assumption in Ref.~\onlinecite{Ivchenko}, an in-plane anisotropic potential is
used in modeling single QDs, i.e.,
$U_{e(h)}=\nu_{e(h)}\theta(b/2-|y_{e(h)}|)\theta(a/2-|x_{e(h)}|)$, where $a$
and $b$ are the lateral sizes of QDs, and $\nu_{e}$ ($\nu_{h}$) is the
conduction (heavy-hole valence) band offset. Whether the geometric shape of
single QDs is rectangular or elliptic will not change the qualitative results
of this letter. $V_{\mathrm{ex}}$ is the electron-hole exchange
interaction.~\cite{Ivchenko} The material parameters used in the calculations
refer to Ref.~\onlinecite{Dong}. The computational procedure is that
eigenfunctions of spin-independent $H_1$ is firstly calculated using
diagonalization method with more than 4000 basis sets, and then FSS is obtained
by calculating the matrix elements of $V_{\mathrm{ex}}$ in the basis of
eigenfunctions of $H_1$.

%In this letter, we consider three different cases, i.e., absence of donor
%impurity ($N=0$), one or two ionized hydrogenic donor impurities ($N=1,2$).

%The bright doublets typically consist of two linearly polarized states, the
%splitting of which (fine-structure splitting) is about tens of $\mu$eV and
%mainly contributed from long-range exchange
%interaction.~\cite{Bayer,Takagahara}

\begin{figure}
\begin{center}
\includegraphics*[angle=0,width=0.45\textwidth]{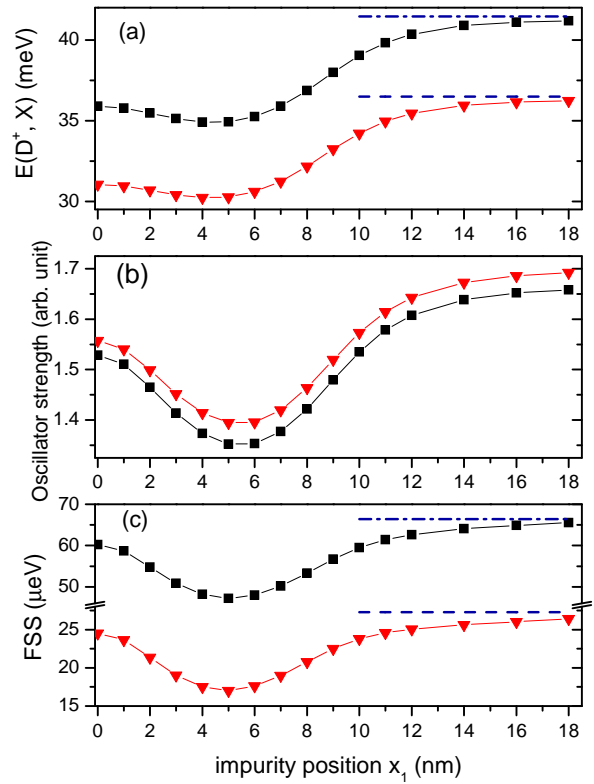}
\caption{(Color online) (a) The ground-state energy, (b) corresponding
oscillator strength , and (c) fine-structure splitting of single ionized donor
impurity-exciton complex as functions of the donor impurity position $x_1$,
with $y_1=0$ for QD1 (triangles) and QD2 (squares), respectively. The exciton
ground-state energy and FSS in QD1 (dash lines) and QD2 (dash-dot lines)
without impurities are also shown, respectively. }\label{FIG:impurity position}
\end{center}
\end{figure}

For simplicity, single ionized donor impurity is located along the $x$-axis,
and the ionized impurity-exciton complex binding energy $E_{b}$ is defined as
\begin{eqnarray}\label{Binding energy}
E_{b}=E(X)-E(D^{+},X),
\end{eqnarray}
where $E(X)$ is the exciton ground state energy in QDs and $E(D^{+},X)$ is the
ground state energy of the ionized donor impurity-exciton complex in the same
QD.~\cite{Stebe} In Fig.~\ref{FIG:impurity position}(a), ground-state energies
of single ionized donor impurity-exciton complex without the electron-hole
exchange interaction in two kinds of anisotropic QDs are shown as functions of
impurity position. For the first kind of QDs (QD1) with $b=18.0$ nm and
$a=20.0$ nm and the donor impurity located at the QD center, $E_b=5.44$ meV,
while for the second kind of QDs (QD2) with $b=16.0$ nm and $a=20.0$ nm and the
donor impurity located at the QD center, $E_b=5.56$ meV. As the impurity
position $x_1$ increases from zero, however, the binding energy initially
increases and gets a maximal value at about $x_1=5.0$ nm for both QD1 and QD2
as shown in Fig.~\ref{FIG:impurity position}(a). When $x_1$ exceeds $5.0$ nm,
$E_b$ becomes smaller and gradually approaches zero as $x_1\rightarrow\infty$.
In Fig.~\ref{FIG:impurity position}(b), corresponding oscillator strength of
the exciton transition is also shown.

Including the exchange interaction $V_{\mathrm{ex}}$, the exciton ground states
are split into bright and dark doublets. FSS of bright doublet is mainly
determined by the anisotropic shape of InGaAs QDs.~\cite{Ivchenko} FSS of QD2
with larger shape anisotropy and without impurity (66 $\mu$eV) is larger than
that of QD1 with smaller shape anisotropy and without impurity (27 $\mu$eV),
and calculated value of FSS is well consistent with recent experimental
results.~\cite{Tartakovskii2} When there is an ionized donor impurity present
at the QD center, FSS is slightly reduced, relative to the $N=0$ case, as shown
in Fig.~\ref{FIG:impurity position}(c). Interestingly, reduction of FSS is
largely enhanced for the off-center donor impurity. At about $x_1=5.0$ nm, FSS
gets a minimal value, i.e., 17 $\mu$eV and 47 $\mu$eV for QD1 and QD2,
respectively. Moreover, FSS could be further reduced when there are more than
one ionized donor impurity. For example, the ground-state FSS of two ionized
donor impurities-exciton complex in QD1 is only 6 $\mu$eV, with the two donor
impurities position $\mathrm{q}_1=(5.0~ \mathrm{ nm},1.0~ \mathrm{ nm})$ and
$\mathrm{q}_2=(5.0~ \mathrm{ nm}, -1.0~ \mathrm{ nm})$.

\begin{figure}
\begin{center}
\includegraphics*[angle=0,width=0.45\textwidth]{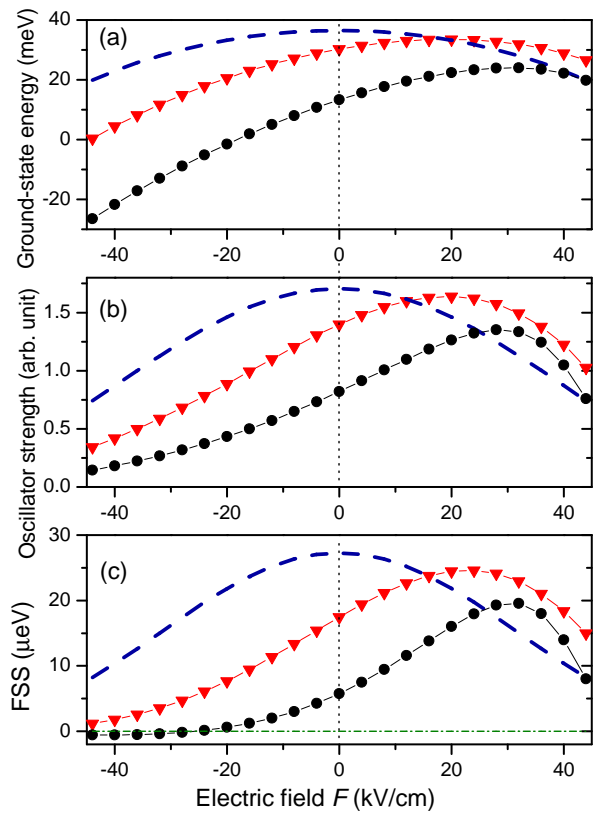}
\caption{(Color online) (a) The ground-state energy, (b) corresponding
oscillator strength, and (c) FSS in QD1 as functions of the external electric
field $F$ for the following three cases: (i) absence of ionized impurity (dash
lines), (ii) one ionized donor impurity (triangles) with $\mathrm{q}_1=(5.0$
nm, $0$ nm), (iii) two ionized donor impurities (circles) with
$\mathrm{q}_1=(5.0$ nm, $1.0$ nm), $\mathrm{q}_2=(5.0$ nm, $-1.0$
nm).}\label{FIG:external electric field}
\end{center}
\end{figure}

% For the same time, the corresponding exciton oscillator strength is not much
% reduced, which is about $50\%$ of that in the same QD without impurity.

In Fig.~\ref{FIG:external electric field}(a), ground-state energies without the
electron-hole exchange interaction for three different systems, i.e., (i)
exciton without impurity, (ii) single ionized donor impurity-exciton complex,
and (iii) two ionized donor impurities-exciton complex in QDs, are shown as
functions of an external electric field along the $x$-axis. The single donor
impurity position is $(5.0~ \mathrm{nm},0 ~\mathrm{nm})$, and the two donor
impurity positions are $(5.0~\mathrm{nm}, 1.0~\mathrm{nm})$ and
$(5.0~\mathrm{nm}, -1.0~\mathrm{nm})$, respectively. The Stark shift of all
three cases could be well approximated by a parametric model
\begin{eqnarray}\label{Stark effect}
\Delta E(F) = \alpha F -\frac{1}{2}\beta F^2+...,
\end{eqnarray}
where $\alpha$ and $\beta$ are actually the exciton dipole and polarizability,
respectively, along the external field.~\cite{Empedocles} For the first case
without ionized impurities, $\alpha$ is zero and the Stark shift could be well
fitted by the quadratic term. However, $\alpha$ becomes nonzero for the second
and third case because of the opposite Coulomb interactions between
electron-ionized donor and hole-ionized donor. At small external electric
field, the Stark shift could be well described by the linear and quadratic
terms, while higher-order terms need to be taken into account at larger
external electric field. Oscillator strength of the ground state shows
interesting behaviors. For the first case, oscillator strength monotonically
decreases as the external electric field, and its behavior is symmetric for the
field in the positive and negative directions. For the last two cases, the
behavior of the ground-state oscillator strength with the external field is
clearly asymmetric. For $F>0$, external field partially counteracts the local
field produced by the ionized impurities, and the overlap between electron and
hole is initially enhanced. Therefore, the ground-state oscillator strength
initially increases. However, as the external field exceeds the effective local
field, then oscillator strength is finally reduced as shown in
Fig.~\ref{FIG:external electric field}(b). For $F<0$, external field is in the
same direction of the effective local field produced by the ionized impurities,
and the oscillator strength monotonically decreases.

FSS of the three cases are shown as functions of the external electric field in
Fig.~\ref{FIG:external electric field}(c). It can be seen that the behavior of
exciton FSS in QDs with off-center ionized impurities is greatly different from
that of QDs without ionized impurities. For QDs without ionized impurities, FSS
monotonically decreases with the external electric field. However, FSS of QDs
with off-center ionized impurities shows asymmetric variations for external
electric field in positive and negative directions. We note that FSS shows
somewhat similar behaviors as functions of the external electric field with
those of the oscillator strength, since both the oscillator strength and FSS
are directly related to the overlap between the electron and hole.
Interestingly, FSS in the third case could be reduced to less than $1$ $\mu$eV
as $F<-16$ kV/cm, which is below the typical homogeneous linewidth of the
exciton emission lines (determined by the exciton radiative lifetime), and the
corresponding oscillator strength at $F=-16$ kV/cm is about $30\%$ of that in
the same QD without ionized impurity and external field. With the help of the
local field produced by the off-center donor impurities, FSS in anisotropic QDs
could be more efficiently tuned towards zero with external electric field as
shown in Fig.~\ref{FIG:external electric field}(c). On the other hand,
according to the analysis mentioned above, it is easy to understand that local
electric field, produced by ionized impurities, defect trapped electron or
hole, and surface charges in real samples could strongly affect the external
electric field-tuning of FSS in QDs. Thus the results in this letter might be
useful to explain the anomalous external electric-field dependence of FSS
observed in the experiment.~\cite{Gerardot}

\begin{figure}
\begin{center}
\includegraphics*[angle=0,width=0.45\textwidth]{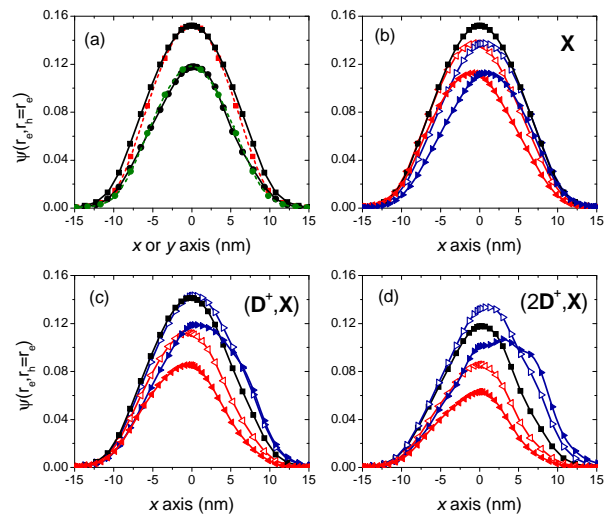}
\caption{(Color online) (a) $\psi(r_e,r_h=r_e)$ along the $x$- (solid lines)
and $y$-axis (dash lines) for the first (squares) and the third case (circles)
of Fig.~\ref{FIG:external electric field}, respectively, with $F=0$ kV/cm;
$\psi(r_e,r_h=r_e)$ along the $x$-axis of (b) the first, (c) the second, and
(d) the third case in Fig.~\ref{FIG:external electric field}, for $F=-40.0$
kV/cm (leftward triangles), $-24.0$ kV/cm (open leftward triangles), 0 kV/cm
(squares), $+24.0$ kV/cm (open rightward triangles), and $+40.0$ kV/cm
(rightward triangles), respectively.}\label{FIG:wavefunction}
\end{center}
\end{figure}

In order to better understand the reduction of exciton FSS in QDs with ionized
impurities, $\psi(r_e,r_h=r_e)$ of the ground state in QD1 along the $x$- and
$y$-axis are shown in Fig.~\ref{FIG:wavefunction}(a). For QD1 without ionized
impurity and external field, the extension of $\psi(r_e,r_h=r_e)$ along the
$x$-axis is slightly larger than that along the $y$-axis as clearly shown in
Fig.~\ref{FIG:wavefunction}(a). Thus the electron-hole long range exchange
interaction is nonzero and FSS is calculated to be 27 $\mu$eV. When there are
ionized donor impurities present as in the third case of Fig.~\ref{FIG:external
electric field}, both the amplitude and shape of $\psi(r_e,r_h=r_e)$ are
greatly changed due to the local electric field produced by the ionized donor
impurities, and the extension along the $x$-axis becomes nearly identical to
that along the $y$-axis as shown in Fig.~\ref{FIG:wavefunction}(a). That is why
FSS is largely reduced to only 6 $\mu$eV. In Figs.~\ref{FIG:wavefunction}(b),
\ref{FIG:wavefunction}(c), and \ref{FIG:wavefunction}(d), $\psi(r_e,r_h=r_e)$
of all the three cases in Fig.~\ref{FIG:external electric field} are shown for
several values of the external electric field, and it could be easily seen that
both the amplitude and shape of $\psi(r_e,r_h=r_e)$ are largely changed by the
external electric field as well as the ionized donor impurities, which clearly
explains the anomalous behaviors of the oscillator strength and FSS in
Figs.~\ref{FIG:external electric field}(b) and \ref{FIG:external electric
field}(c).

In summary, we study the ground-state FSS of ionized shallow donor
impurities-exciton complex in anisotropic QDs, and find that it could be
largely reduced by one or two off-center ionized donor impurities, which
strongly influence the exciton envelope functions. Then anomalous Stark shifts
and efficient tuning of FSS by the external electric field are clearly shown
and discussed. The study will be helpful and interesting for the research on
the QDs-based entangled-photon source.

Financial supports from NSF-China (Grants No. 10574077 and 10774085), the
``863" Programme (No. 2006AA03Z0404) and MOST Programme (No. 2006CB0L0601) of
China are gratefully acknowledged.

\newpage

%\begin{references} \

\newpage

\end{document}